# Power Law Distributions in Korean Household Incomes

Kyungsik Kim* and Seong-Min Yoon†

*Department of Physics, Pukyong National University, Pusan 608-737, Korea*

†*Division of Economics, Pukyong National University, Pusan 608-737, Korea*

ABSTRACT

We investigate the distribution function and the cumulative probability for Korean household incomes, i.e., the current, labor, and property incomes. For our case, the distribution functions are consistent with a power law. It is also showed that the probability density of income growth rates almost has the form of a exponential function. Our obtained results are compared with those of other numerical calculations.

*Corresponding author. Tel.: +82-51-620-6354; fax: +82-51-611-6357.
E-mail address: kskim@pknu.ac.kr.

Recently, the investigation of scaling relations in the Zipf's and Pareto's laws has received considerable attention on econophysics. Until present, there has been mainly concentrated on the distribution of the personal income, the distribution of the company size and income, the scaling relation of company's size fluctuations and the disribution of the city size. In present, these studies are both of much interest and of practical relevance in estimating the dynamical behavior of statistical quantities in financial markets. From the result of these studies, there has led to a better insight for understanding scaling properties, particularly on the basis of novel statistical methods and approaches of economics.

On the other hand, more than one hundred years ago, Pareto [1] investigated the income and wealth distributions described by the characteristic feature of a national economics. Gini also showed that the income distribution is approximated by a power law with the non-universal exponent [2]. Mandelbrot [3] reported that the distribution of incomes scales as a power law, and Montroll and Schlesinger [4] showed that the income of high-income group follows a power law while that of low-income group follows a log-normal distribution [5]. Stanley et al. [6] recently found that American company size distribution is closer to the log-normal law. Okuyama and Takayasu [7] analyzed Japanese and international company databases and reported that the distribution function of annual income of companies shows a power law distribution consistent with the Zipf's law. Furthermore, many researchers have attempted to compute the distribution functions in scientific fields such as the voting process [8], the firm bankruptcy [9], the investment strategy [10], the expressed genes [11, 12], and the earthquake [13].

As the first step, it may be of importance to elucidate the scaling properties and statistical methods for Korean household incomes. Our purpose in this paper is to find the scaling relation in three distributions; the distributions of the current, labor, and property incomes. We also calculate the cumulative probability and the probability density of income growth rates for Korean household incomes.

In this study, we first analyze the database of annual household incomes of 21,132 households for the one-year period of 2000 in Korea. The income distribution is represented in terms of

$$I_i \propto R^{-\alpha_i}, \qquad (1)$$

where $i = C, L$, and $P$ denote the current, labor, and property incomes with scaling exponents $\alpha_C$, $\alpha_L$, and $\alpha_P$, respectively, and $R$ denotes the rank of income groups. When $\alpha_i = 1$, Eq. (1) is satisfied with the Zipf's law. Fig. 1 shows the distribution of the current income in Korea, and the least-squares fit gives a power law with exponent $\alpha_C = -0.31$. From our database, we obtain that the estimated slopes of the labor and property incomes are respectively, $\alpha_L = -0.22$ and $\alpha_p = -0.65$, inconsistent with the Zipf's law, as plotted in Figs. 2 and 3. However, we find from our result that the high-income group earns more property incomes than the low-income group and that the inequality of the current income among households is actually due to that of the property income.

Next, the cumulative probability of an income is defined as

$$P(x \geq I_i) \propto x^{-\beta_i}, \qquad (2)$$

where $β_c$, $β_L$, and $β_p$ are, respectively, the scaling exponents for the cumulative probability of the current, labor, and property incomes. In Fig. 4, we plot the cumulative probability of the current income in Korea, where the slope of the dot line is $β_c = -3.06$. It is found from Figs. 5 and 6 that cumulative probabilities of the labor and property incomes follows a power law with scaling exponents $β_L = -4.78$ and $β_p = -2.15$, respectively.

Lastly, let $s(t)$ be an averaged current income at year $t$. We can show the ratio of the current incomes of successive years as follows;

$$r = \log \frac{s(t+1)}{s(t)} \ . \qquad (3)$$

We analyze the averaged current incomes by job categories for 29 years (from 1965 to 2003) from database of Korea National Statistical Office, and Fig. 7 depicts the probability density of current income growth rate as a function of the normalized current income $r^* = (r - <r>)/σ$, where $<r>$ and $σ$ denote, respectively, the mean value and the standard deviation of $r$. It is also found that the probability density of income growth rates almost has the form of an exponential function, which is proportional to $\exp(-λr^*)$ with $λ=0.5$. It is really known that Japanese income distribution [5] is closer to Zipf's law, while a power law holds in the case of USA [4].

In conclusions, we have found the distribution function, the cumulative probability, and the probability density of income growth rates for Korean household incomes. Particularly, it may be expected from Figs. 1-3 that the high-income group earn more property incomes than the low-income group. The cumulative probability of the current, labor, and property incomes are consistent with a power law, significantly different from Zipf's law in Japanese income distribution [5]. It is also showed that the probability density of income growth rates almost scales as an exponential function. In future, we will extensively investigate the tick data of stock prices in Korean financial markets and compare in detail with calculations performed in other nations.

# Figure Captions

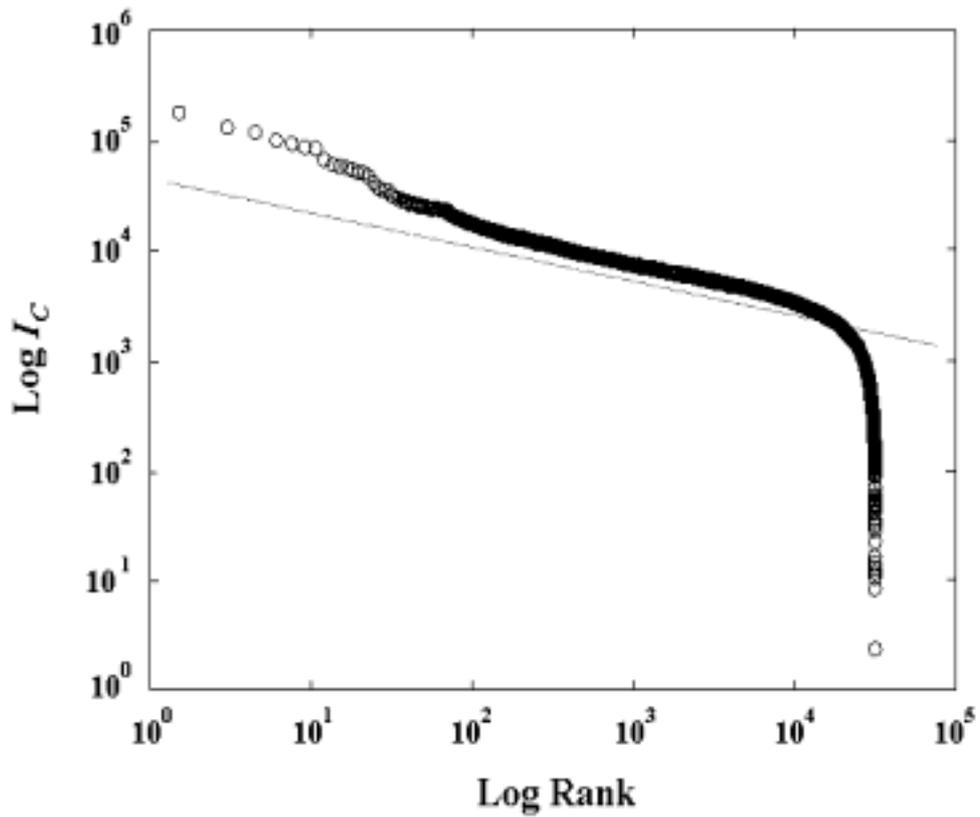

**Fig. 1** Distribution of the current income, where the least-squares fit gives a power law with exponent $\alpha_c = -0.31$.

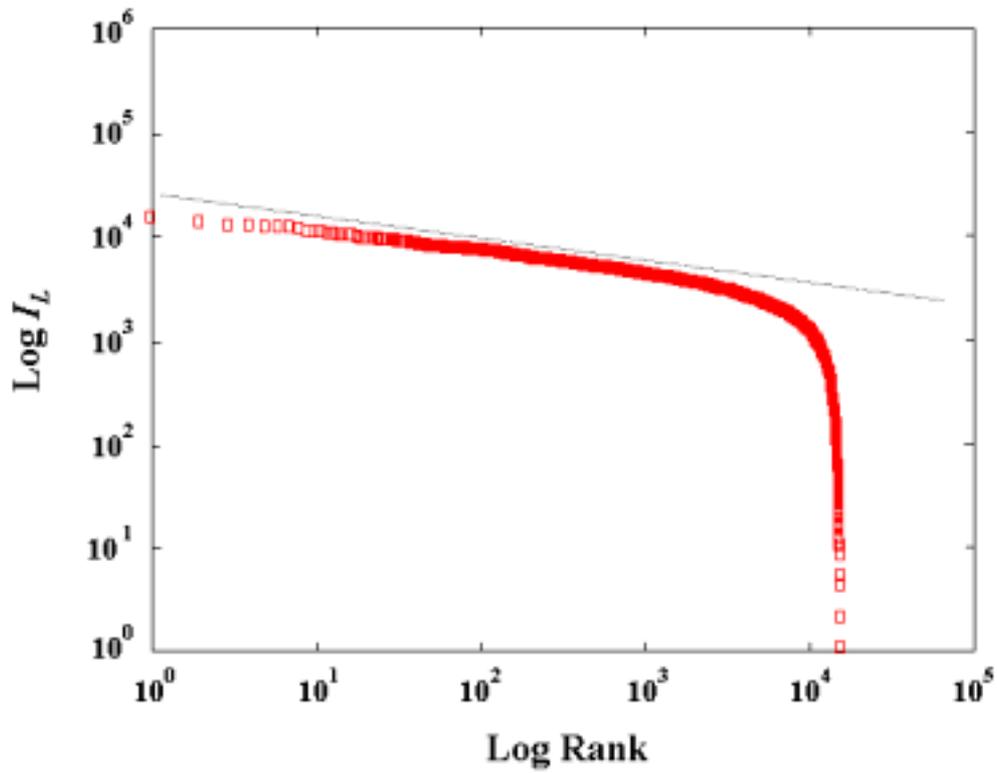

**Fig. 2** Plot of distribution of the labor income scaled as a power law with a scaling exponent $\alpha_L = -0.22$.

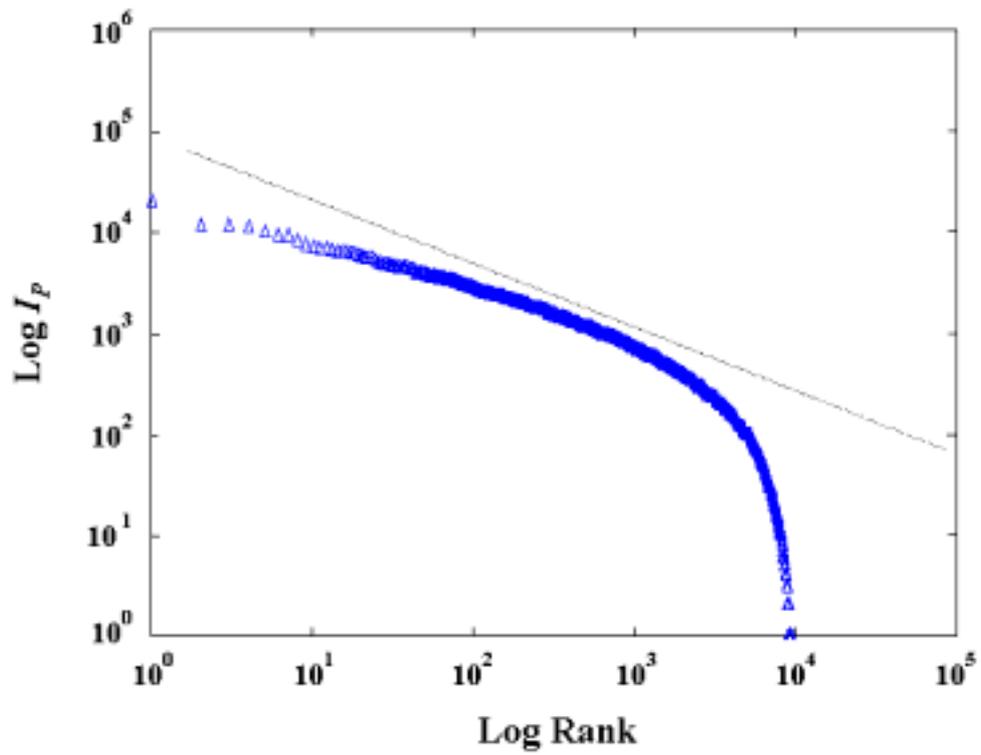

**Fig. 3** Plot of distribution of the property income scaled as a power law with a scaling exponent $\alpha_p = -0.65$.

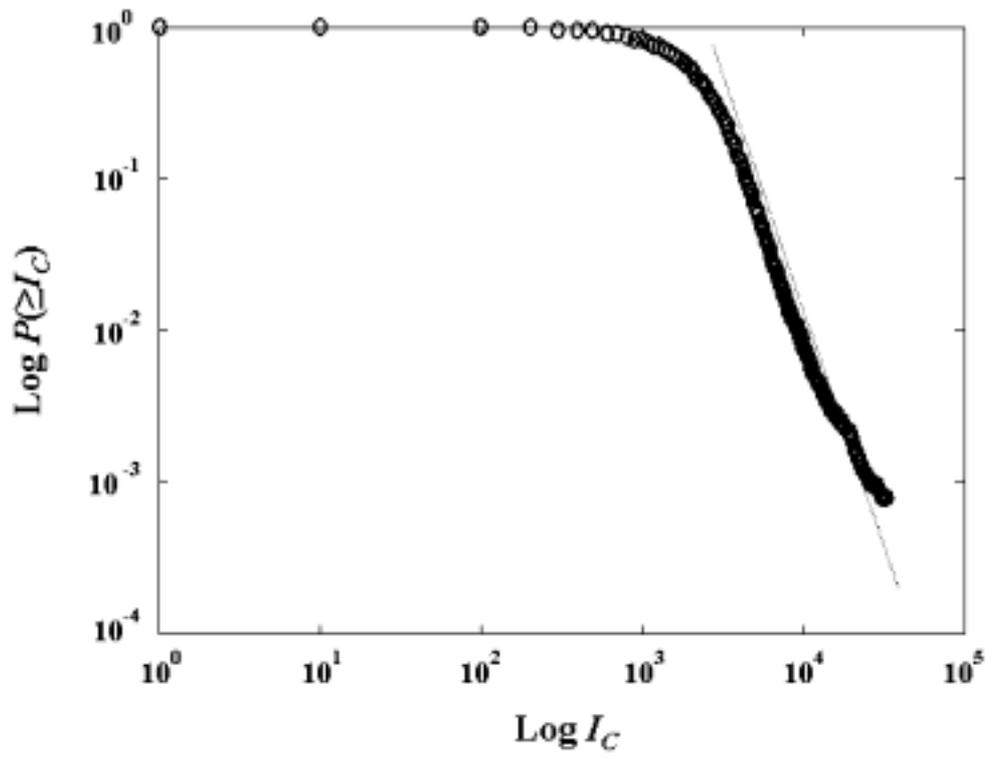

**Fig. 4** Plot of cumulative probability of the current income, where the slope of the dot line is $\beta_c = -3.06$.

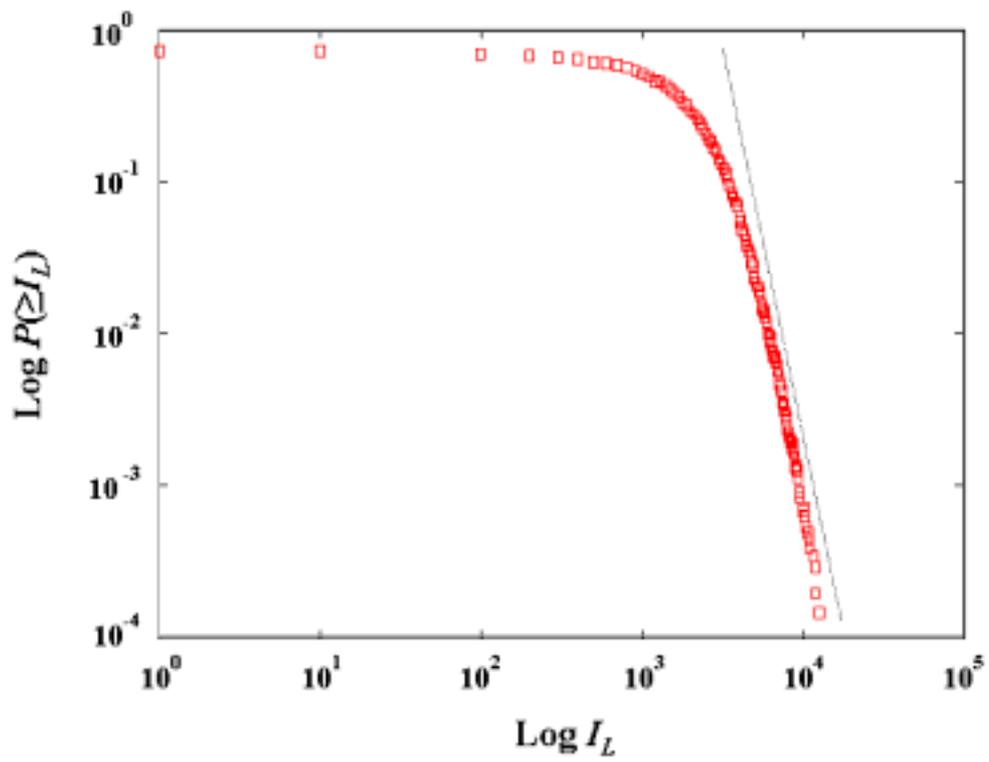

**Fig. 5** Plot of cumulative probability of the labor income, where the slope of the dot line is $\beta_L = -4.78$.

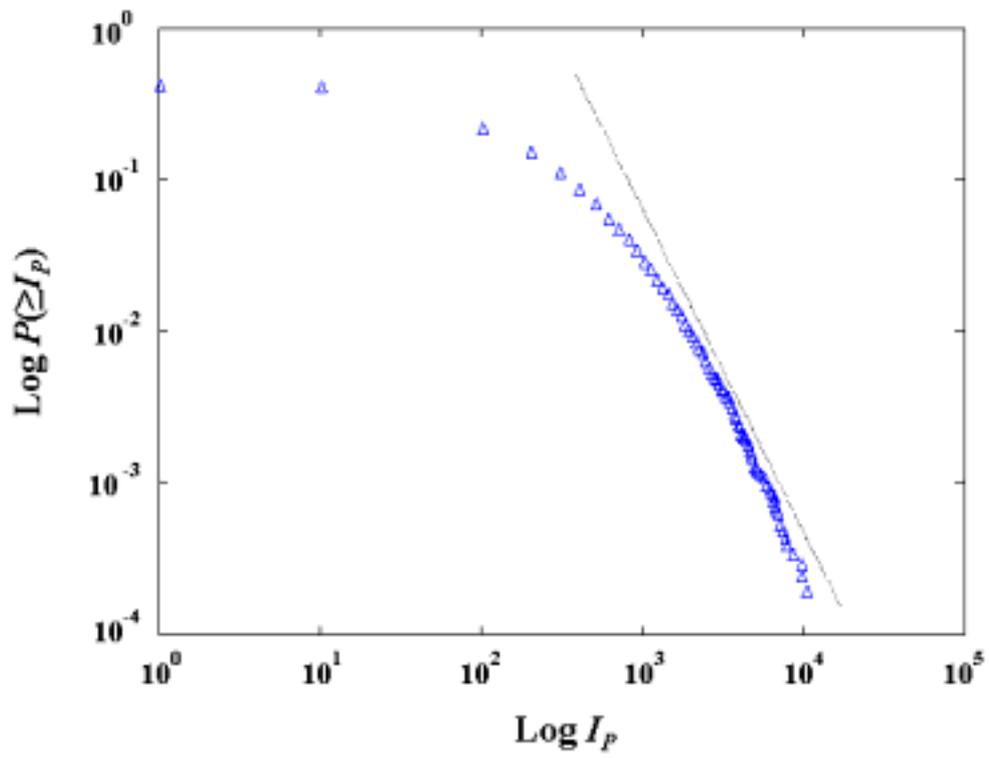

**Fig. 6** Plot of cumulative probability of the property income, where the slope of the dot line is $\beta_p = -2.15$.

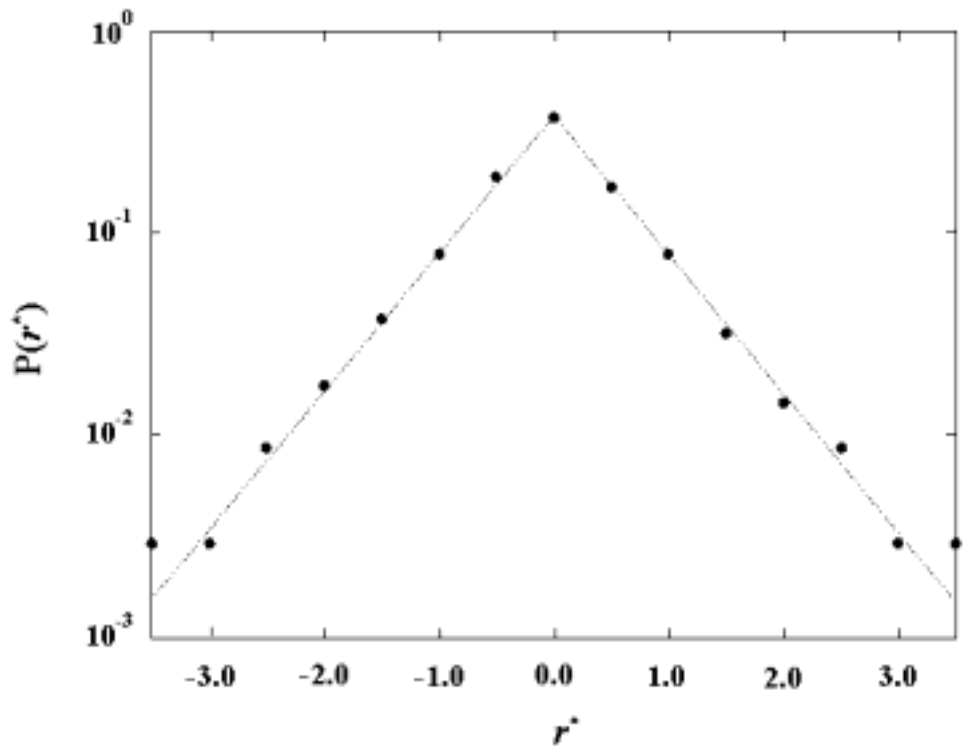

**Fig. 7** Plot of probability density of current income growth rate as a function of the normalized current income $r^*$, where $r^* = (r - <r>)/\sigma$, $<r> = 14.5$, and $\sigma = 14.6$.